\documentclass[%
 aip,
 amsmath,amssymb,
 reprint,%
]{revtex4-1}

\usepackage{graphicx}
\usepackage{dcolumn}
\usepackage{bm}

\usepackage[utf8]{inputenc}
\usepackage[T1]{fontenc}
\usepackage{mathptmx}
\usepackage{etoolbox}
\usepackage{comment}
\usepackage{appendix}
\usepackage{enumitem}
\usepackage{mathtools}
\usepackage[dvipsnames]{xcolor}
\usepackage[colorlinks=true,
            linkcolor=cyan,
            citecolor=cyan,
            urlcolor=cyan]{hyperref}

\makeatletter
\def\@email#1#2{%
 \endgroup
 \patchcmd{\titleblock@produce}
  {\frontmatter@RRAPformat}
  {\frontmatter@RRAPformat{\produce@RRAP{*#1\href{mailto:#2}{#2}}}\frontmatter@RRAPformat}
  {}{}
}%
\makeatother
\begin{document}

\preprint{AIP/123-QED}

\title[]{Analytical solution for dynamic evaporation of liquid in isothermal condition}
\author{Luiz Eduardo Czelusniak}
\email{luiz.czelusniak@partner.kit.edu}
\affiliation{
Lattice Boltzmann Research Group (LBRG), Karlsruhe Institute of Technology (KIT), 76131 Karlsruhe, Germany
}
\affiliation{
Institute for Applied and Numerical Mathematics (IANM), Karlsruhe Institute of Technology (KIT), 76131 Karlsruhe, Germany
}%

\author{Tim Niklas Bingert}
\affiliation{
Lattice Boltzmann Research Group (LBRG), Karlsruhe Institute of Technology (KIT), 76131 Karlsruhe, Germany
}
\affiliation{ 
Institute of Mechanical Process Engineering (MVM), Karlsruhe Institute of Technology (KIT), 76131 Karlsruhe, Germany
}%

\author{Stephan Simonis}
\affiliation{ 
Institute for Applied and Numerical Mathematics (IANM), Karlsruhe Institute of Technology (KIT), 76131 Karlsruhe, Germany
}%
\affiliation{
Lattice Boltzmann Research Group (LBRG), Karlsruhe Institute of Technology (KIT), 76131 Karlsruhe, Germany
}

\author{Alexander J. Wagner}%
\affiliation{ 
Department of Physics, North Dakota State University (NDSU), Fargo, ND 58102, United States
}%

\author{Mathias J. Krause}
\affiliation{
Lattice Boltzmann Research Group (LBRG), Karlsruhe Institute of Technology (KIT), 76131 Karlsruhe, Germany
}
\affiliation{ 
Institute for Applied and Numerical Mathematics (IANM), Karlsruhe Institute of Technology (KIT), 76131 Karlsruhe, Germany
}%
\affiliation{ 
Institute of Mechanical Process Engineering (MVM), Karlsruhe Institute of Technology (KIT), 76131 Karlsruhe, Germany
}%

\date{\today}

\begin{abstract}
An analytical solution based on a diffuse interface model is presented for an isothermal evaporation problem under sub-saturation pressure. The macroscopic equations are derived from the free-energy method, widely recognized in the lattice Boltzmann literature, distinguishing our approach from conventional evaporation models that rely on jump conditions or pure kinetic theory. The interface behavior is fully described by differential equations, eliminating the need for assumptions such as local equilibrium at the interface. We derive an exact analytical solution for the inviscid case and propose an approximate solution when viscosity effects are considered. Our model unveils a novel relationship between evaporation rate and viscosity, providing new insights that have not been thoroughly explored in the literature. The analytical results are validated through numerical simulations using the open-source parallel library \textit{OpenLB}, demonstrating excellent agreement in predicting the physical behavior of the evaporation phenomena within the framework of diffuse interface methods.
\end{abstract}

\maketitle



\section{\label{sec:level1}Introduction}

Understanding and predicting the evaporation phenomena is crucial for engineering applications, e.g. food drying\cite{fathi2022emerging}, industrial dehydration\cite{al2023multi}, distillation process\cite{skiborowski2023synthesis}, vacuum evaporation\cite{abdelrahman2025vacuum}, paint drying\cite{di2023dynamics}, evaporative cooling\cite{abdullah2023technological}, steam power generation\cite{polski2024novel}, and others.  
There are different ways to study this phenomenon: analytically\cite{hertz1882ueber,knudsen1909gesetze,holyst2015molecular,persad2016expressions,schrage1953theoretical,stefan1891theorie,scriven1995dynamics,cossali2023analytical,alvarez2024fully}, numerically\cite{municchi2022conjugate,wang2023pool,municchi2024computational}, and experimentally\cite{manova2022experimental,baptistella2023liquid,marchetto2024experimental}. For highly complex problems, numerical and experimental approaches are essential. However, the analytical method provides physical insights that lead to a deeper understanding of the phenomenon, making it highly valuable.

There is extensive literature on evaporation modeling. An important result is the Hertz--Knudsen relation~\cite{hertz1882ueber,knudsen1909gesetze,holyst2015molecular,persad2016expressions}. Another significant contribution involving kinetic theory was made by Schrage~\cite{schrage1953theoretical}, who calculated evaporation fluxes of pure substances and multi-component systems. Notable results employing macroscopic phase change modeling include the one-dimensional Stefan problem~\cite{stefan1891theorie}, originally developed for solid--liquid phase transition and Scriven's solution~\cite{scriven1995dynamics} for droplet evaporation. Recent articles extend these approaches to more complex situations~\cite{cossali2023analytical,alvarez2024fully}. 

Another approach to modeling multiphase systems that has gained popularity in numerical studies is the diffuse interface model. This model treats the interface as a thin region in which fluid properties continuously transition between the vapor and liquid phases. The phase-field model~\cite{jacqmin1999calculation,lamorgese2011phase,kim2012phase}, which usually employs the Cahn--Hilliard~\cite{novick2008cahn,lee2014physical} or the Allen--Cahn~\cite{feng2003numerical} equation, is a diffuse interface model used in combination with several numerical methods, for example: lattice Boltzmann methods (LBM)~\cite{wang2016comparative}, finite element methods (FEM)~\cite{feng2007analysis}, finite volume (FV)~\cite{qiao2015phase} or immersed boundary methods (IBM)~\cite{hua2014level}. Other examples of diffuse interface model-based methods are the free-energy~\cite{swift1996lattice,semprebon2016ternary,simonis2024binary} and the pseudo-potential~\cite{shan1993lattice,shan1994simulation} LBM, which are single-component methods capable of modeling liquid-vapor phase transition without any special treatment of the interface. 
Due to good parallelizability and robustness, LBM are generally well-suited for complex flow applications~\cite{ito2024identification,bukreev2023consistent,bukreev2024benchmark,mink2022comprehensive}, even at large scales~\cite{kummerlander2024advances}. 
Notably, many of the diffuse interface models listed above (or derivates thereof) are implemented in open-source libraries such as \textit{OpenLB}~\cite{openlb2020} and thus readily available for parallel execution on personal devices or clusters of any size.

While diffuse interface models such as Cahn–Hilliard are well-established in numerical and analytical studies of phase separation\cite{foard2009enslaved}, their use in analytical descriptions of liquid evaporation processes remains limited. However, these models also hold great potential for the development of analytical solutions since the fluid behavior is fully described by differential equations. This eliminates the need for assumptions such as local equilibrium at the interface, which is commonly used in sharp interface modeling. This represents a significant advancement in the analysis of non-equilibrium thermodynamics.

In this work, we employ a diffuse interface model to study the evaporation process. To simplify the problem, we consider an isothermal system subject to a sub--saturation pressure condition. A similar system was previously studied by Jamet~\cite{jamet2004test}, but using a sharp interface approach. With our innovative methodology, we observed a dependence of the evaporation rate on the fluid viscosity -- a behavior that has not been captured by any existing model in the literature. Finally, we derived an analytical solution for the evaporation rate in the inviscid case and an approximate solution for the viscous case. The predictions from our solutions are confirmed by simulations using an LBM implemented in the open source parallel library \textit{OpenLB}~\cite{openlb2020}.

The practical applications of our analytical solution are as follows: Firstly, it provides predictions that can be tested in experiments to validate diffuse interface methods or suggest future corrections to this approach. Secondly, our solution can also be used to construct benchmark cases to test numerical approaches for diffuse interface models. Thirdly: predicting evaporation phenomena under sub-saturation pressure is crucial in, e.g., fuel injection systems, especially within low-pressure chambers. This process has significant implications for combustion engines\cite{zeng2012atomization}. 
In future works, our approach of constructing an analytical solution can be extended to non--isothermal and multi--component problems.

The article is organized as follows: In 
Section~\ref{sec:Problem}, the physical problem of isothermal evaporation under sub-saturation pressure is introduced. In Section~\ref{sec:Equations}, the governing equations are presented. Section~\ref{sec:Solution} details the analytical solution procedure, including the derivation of the interface velocity and its dependence on physical parameters. In Section~\ref{sec:Results}, the analytical solution is evidenced through comparisons with LBM simulations, and the influence of viscosity and pressure differences on the interface dynamics is analyzed. Finally, in Section~\ref{sec:Conclusion}, we summarize and discuss the main findings and propose directions for future work.



\section{\label{sec:Problem}Problem description}

The problem of interest in this work is a dynamic isothermal liquid-vapor evaporation process similar to the one described by Jamet~\cite{jamet2004test}. We consider a system with two phases (one liquid and one vapor) separated by a flat interface, which reduces to a one-dimensional case. The system has a fixed temperature $T$. In equilibrium, this system would have a homogeneous and constant pressure called saturation pressure $p_{\mathrm{sat}}(T)$. Our interest is to study what happens in this system when a pressure smaller than the saturation pressure is imposed in the vapor phase.

\begin{figure}[ht] 
\centering
	\includegraphics[width=\columnwidth]{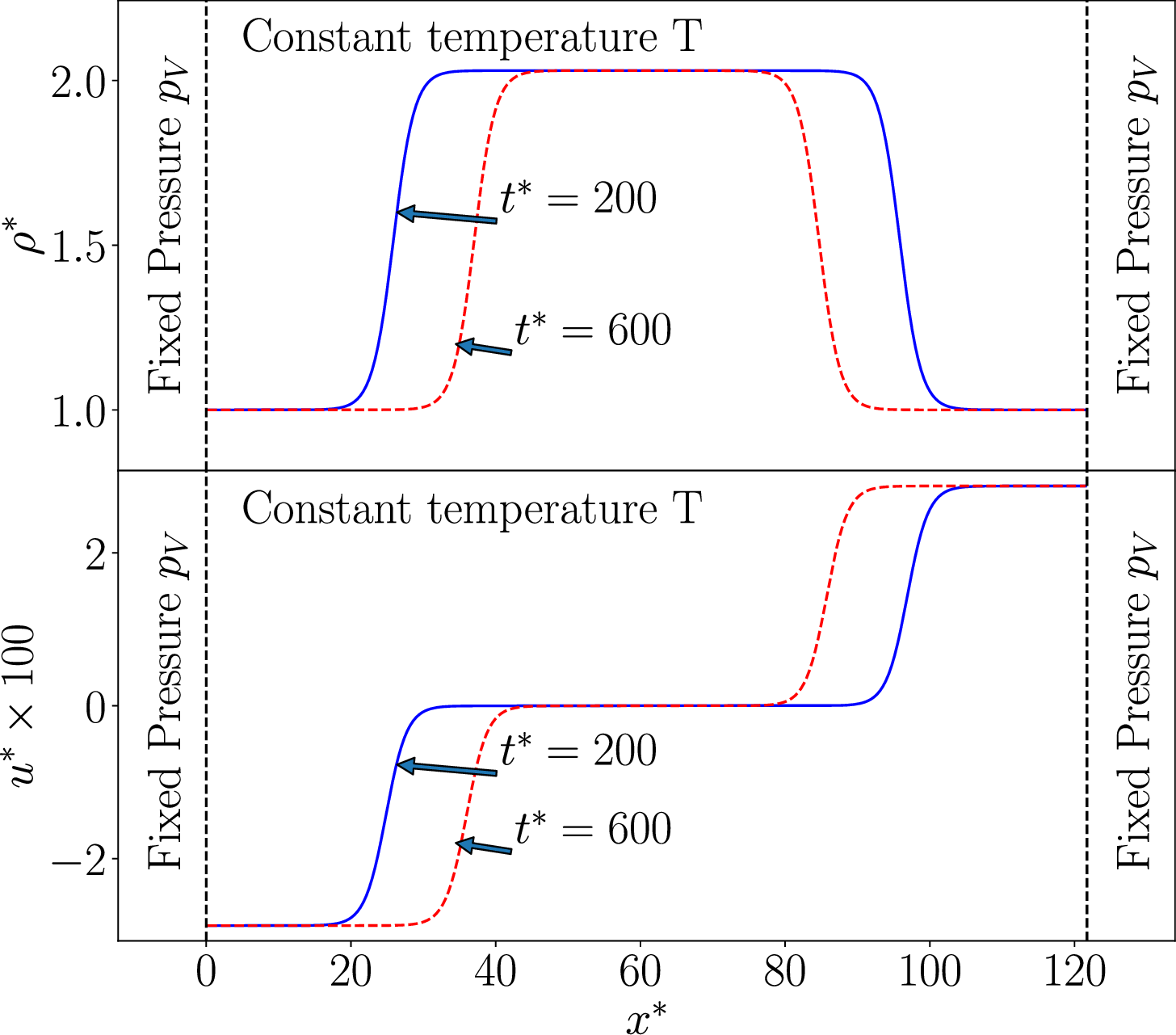}
	\caption{Top: density $\rho^{\ast}$ along domain size $x^{\ast}$ at two instants $t^{\ast}$. Bottom: velocity $u^{\ast}$ for the same conditions. Dimensionless quantities defined in \eqref{eq:dimensionless}. 
    Simulations conducted with the model of Wagner~\cite{wagner2006thermodynamic} implemented in \textit{OpenLB}~\cite{openlb2020} (see Section~\ref{sec:Numerical} for details).}
    \label{fig:ProblemDescription}
\end{figure}

To illustrate this phenomenon, we use a simulation performed with the free-energy LBM formulation proposed by Wagner~\cite{wagner2006thermodynamic}. Figure~\ref{fig:ProblemDescription}
shows a representation of the simulated system. The top plot shows the density profile $\rho^{\ast}$ at two instants of time ($t^{\ast}=200$ and $t^{\ast}=600$). On the left and right boundaries (vapor phase) a pressure lower than the saturation pressure $p_{\text{v}}=0.99p_{\text{sat}}$ is imposed.
The superscript $\cdot^{\ast}$ indicates dimensionless quantities such as $t^{\ast}$. The definition of the dimensionless quantities is presented in Section~\ref{sec:Equations}. The simulation details are presented in Section~\ref{sec:Numerical}. 

We observe in the density profile of Figure~\ref{fig:ProblemDescription} that the liquid region shrank over time, which means that the imposed boundary condition led to an evaporation process. The lower plot of Figure~\ref{fig:ProblemDescription} shows the velocity profile. In the liquid region, the velocity is zero due to the problem symmetry. However, the velocity profile indicates that mass is leaving the domain through both boundaries.

\begin{figure}[ht] 
\centering
	\includegraphics[width=\columnwidth]{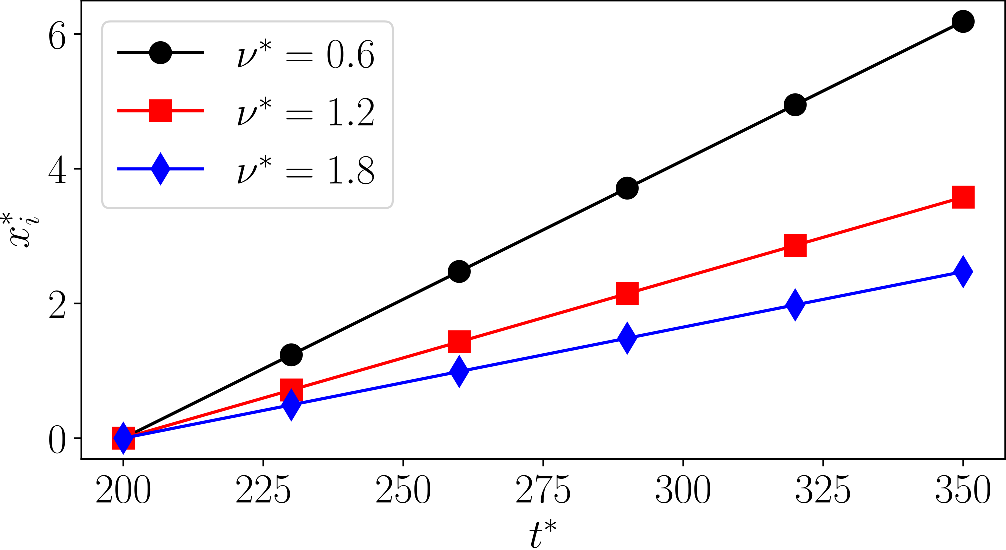}
	\caption{Interface position $x_i^{\ast}$ over time $t^{\ast}$ for several viscosities $\nu^{\ast}$. Dimensionless quantities described in~\eqref{eq:dimensionless}. Simulations conducted with the model of Wagner~\cite{wagner2006thermodynamic} implemented in \textit{OpenLB}~\cite{openlb2020} (see Section~\ref{sec:Numerical} for details).}
    \label{fig:MovingInterface}
\end{figure}

This simulation was performed for a uniform dimensionless viscosity $\nu^{\ast}=1$. We repeated the simulation for different viscosity values and recorded the interface position for several other time instants. The results are summarized in Figure~\ref{fig:MovingInterface}. Two very interesting behaviors are observed. First, for sufficiently long simulation times, the "position vs.\ time" curve is a straight line, which implies a constant interface velocity. The second surprising result is that the interface velocity depends on the fluid viscosity. This dependence of the evaporation rate on fluid viscosity is not present in classical evaporation models\cite{hertz1882ueber,knudsen1909gesetze,schrage1953theoretical} nor in analytical solutions to evaporation problems\cite{stefan1891theorie,scriven1995dynamics}. Showing, to the best of our knowledge, a behavior that has not yet been described in the literature.

In the next sections, we focus on obtaining an analytical solution that describes this physical phenomenon.



\section{\label{sec:Equations}Governing equations}

This problem is modeled by the one-dimensional (1D) mass and momentum  conservation equations:
\begin{subequations}
\begin{equation}\label{eq:mass}
\frac{ \partial \rho }{ \partial t } 
+ \frac{ \partial (\rho u) }{ \partial x } = 0,
\end{equation}
\begin{equation}\label{eq:momentum}
\rho \frac{ \partial u }{ \partial t }
+ \rho u \frac{ \partial u }{ \partial x } 
= - \frac{ \partial p }{ \partial x } 
+ \frac{ \partial \tau_{xx} }{ \partial x },
\end{equation}
\end{subequations}
where \(\rho\) denotes the density, \(u\) is the velocity, \(p\) is the pressure, and $\tau_{xx}$ is the viscous stress as presented in Kr\"{u}ger \textit{et al.}\cite[p. 6]{kruger2017lattice}:
\begin{equation} \label{eq:stress_complete}
\tau_{xx} = \left( \frac{4}{3} \nu + \nu_{\text{b}} \right) \rho \frac{ \partial u }{ \partial x }, 
\end{equation}
and, unless stated otherwise, all variables depend on time \(t\) and space \(x\). 
The symbols $\nu>0$ and $\nu_{\text{b}}$ in \eqref{eq:stress_complete} represent the shear and bulk kinematic viscosity coefficients, respectively.
For dilute monatomic gases, kinetic theory~\cite{vincenti1966introduction} predicts $\nu_{\text{b}}=0$ which coincides with Stokes' hypothesis~\cite{stokes2007theories}. Notably, the bulk viscosity for non-ideal fluids is a complex topic~\cite{graves1999bulk,kosuge2022navier}. In addition, the stress-tensor behavior at the interface is barely described in the literature. 

In particular, in this work, we compare the analytical solution of a problem governed by \eqref{eq:mass} and \eqref{eq:momentum} with the numerical simulations performed with LBM based on an isothermal flow assumption. 
Thus, we use a different stress tensor definition than \eqref{eq:stress_complete}, i.e.\
\begin{equation} \label{eq:stress}
\tau_{xx} = 2 \nu \rho \frac{ \partial u }{ \partial x }. 
\end{equation}
The alternative form of the stress tensor given in Eq.\eqref{eq:stress}, compared to the complete expression in \eqref{eq:stress_complete}, arises from the fact that the bulk viscosity predicted by the Bhatnagar--Gross--Krook~\cite{bhatnagar1954model}--based lattice Boltzmann model is incompatible with the assumption of isothermal flow. Unless stated otherwise, we continue with the stress tensor defined in \eqref{eq:stress}. The implications of this consideration can be tested, and if necessary, future corrections can be made. 

Besides, we use a Korteweg-type pressure\cite{korteweg1901forme}
\begin{equation} \label{eq:Korteweg}
p = p_{\mathrm{EOS}}
+ \frac{\kappa}{2} \left( \frac{ \partial \rho }{ \partial x } \right)^2
- \kappa \rho \frac{ \partial^2 \rho }{ \partial x^2 },
\end{equation}
where $p_{\text{EOS}}$ denotes the equation of state (EOS) and $\kappa>0$ is the surface tension coefficient.

A fixed pressure $p_{\text{v}}$ is imposed on the boundary. Since the temperature is fixed, this will imply a fixed vapor density at the boundary $\rho_{\text{v}}$. Then, we define the following dimensionless quantities:
\begin{equation} \label{eq:dimensionless}
\begin{aligned}
x^{\ast} &\coloneqq \frac{x}{\rho_{\text{v}}}
\sqrt{\frac{p_{\text{v}}}{\kappa}}, 
~~~ t^{\ast} \coloneqq t\frac{p_{\text{v}}}{\sqrt{\rho_{\text{v}}^3\kappa}},
~~~ u^{\ast} \coloneqq 
u\sqrt{\frac{\rho_{\text{v}}}{p_{\text{v}}}}, 
\\
\nu^{\ast} & \coloneqq \frac{\nu}{\sqrt{\rho_{\text{v}}\kappa}}, 
~~~~~ \rho^{\ast} \coloneqq \frac{\rho}{\rho_{\text{v}}}, 
~~~~~~~~~~~ p^{\ast} \coloneqq \frac{p}{p_{\text{v}}}.
\end{aligned}
\end{equation}
%



\section{\label{sec:Solution}Ansatz for a model solution}


When performing initial simulations, we observed that the interface was moving with a constant speed (cf.\ Figure~\ref{fig:MovingInterface}). Thus, in the ansatz used further below to derive an analytical solution, we make explicit use of the following assumptions:
\begin{enumerate}
    \item After a transient time, the solution reaches a state where the interface propagates with constant velocity $u_{\text{int}}$.
    \item The center of the drop will have a zero velocity by symmetry and will retain a constant density.
    \item The previous assumptions suggest that we can pick a frame of reference ($x',t'$)  for half of the system that moves with the interface, where any field $\phi$ is then independent of time, i.e.\ $\partial\phi/\partial t'=0$.
\end{enumerate}

The transformation between the old and new frame of reference is: 
\begin{equation}
\begin{aligned}
x' &= x - u_{\text{int}} t, \\
t' &= t, 
\end{aligned}
~~~~~~~~~~
\begin{aligned}
x &= x'  + u_{\text{int}} t', \\
t &= t'.
\end{aligned}
\end{equation}

From the definitions in \eqref{eq:dimensionless} and the consideration of Assumption~3, it follows that:
\begin{equation}
\begin{aligned}
\frac{\partial \phi}{\partial t} &=
\left( \frac{\partial \phi}{\partial t'} \right)
\frac{\partial t'}{\partial t}
+
\left( \frac{\partial \phi}{\partial x'} \right)
\frac{\partial x'}{\partial t}
\quad\Rightarrow\quad
\frac{\partial \phi}{\partial t} =
- u_{\text{int}} \frac{\partial \phi}{\partial x'} , \\
\frac{\partial \phi}{\partial x} &=
\left( \frac{\partial \phi}{\partial t'} \right)
\frac{\partial t'}{\partial x}
+
\left( \frac{\partial \phi}{\partial x'} \right)
\frac{\partial x'}{\partial x}
\quad\Rightarrow\quad
\frac{\partial \phi}{\partial x} =
\frac{\partial \phi}{\partial x'} .
\end{aligned}
\end{equation}

\subsection{Mass conservation}

Now, the mass conservation equation \eqref{eq:mass} is rewritten in terms of $x'$. Since the solution is independent of $t'$, we write $\partial\phi/\partial x'=d\phi/dx'$ which gives
\begin{equation}\label{eq:mass_rewritten}
- u_{\text{int}} \frac{d\rho}{dx'} + \frac{ d (\rho u) }{ dx' } = 0
\quad\Rightarrow\quad
( u_{\text{int}} - u )\frac{ d\rho }{ dx' } = \rho \frac{ du }{ dx' }.
\end{equation}
We integrate \eqref{eq:mass_rewritten} and obtain a relation between $\rho$ and $u$ up to a constant \(C_{1}\):
\begin{equation} \label{eq:RhoUequation}
\int \frac{ d\rho }{ \rho } 
= \int \frac{ du }{ u_{\text{int}} - u } + C_1 
\quad\Rightarrow\quad
\rho ( u_{\text{int}} - u ) = \mathrm{e}^{C_1} = \text{const} .
\end{equation}
According to Assumpion~2, in the liquid region $\rho=\rho_{\text{l}}$ and $u=0$, respectively. Then, from \eqref{eq:RhoUequation} we obtain
\begin{equation} \label{eq:RhoUrelation}
    \rho(u_{\text{int}}-u)=\rho_{\text{l}}u_{\text{int}}.
\end{equation}
Thus, our main finding from the mass conservation equation \eqref{eq:mass} is \eqref{eq:RhoUrelation} that leads to the useful relation:
\begin{equation} \label{eq:RhoUderivative}
\frac{du}{dx'} = \frac{(u_{\text{int}}-u)}{\rho}\frac{d\rho}{dx'}
\quad\Rightarrow\quad
\frac{du}{dx'} = \frac{\rho_{\text{l}} u_{\text{int}}}{\rho^2}\frac{d\rho}{dx'}
\end{equation}
We make explicit use of \eqref{eq:RhoUderivative} below. 

\subsection{Momentum conservation}

Considering the momentum equation~\eqref{eq:momentum}, we start by replacing the derivatives in terms of $t$ and $x$ with derivatives in terms of $x'$, which gives
\begin{equation}\label{eq:momentum_replacedDerivatives}
- \rho u_{\text{int}} \frac{ d u }{ dx' } 
+ \rho u \frac{ d u }{ dx' } 
= - \frac{ dp }{ dx' }
+ \frac{ d }{ dx' }
\left( 2 \nu \rho \frac{ du }{ dx' } \right).
\end{equation}
Next, we rewrite the left-hand side of \eqref{eq:momentum_replacedDerivatives} using \eqref{eq:RhoUrelation} and \eqref{eq:RhoUderivative} to obtain
\begin{equation}\label{eq:momentum_LeftHandRho}
\rho (u-u_{\text{int}}) \frac{ d u }{ dx' }
= -\frac{\rho_{\text{l}}^2u_{\text{int}}^2}{\rho^2}\frac{d\rho}{dx'} .
\end{equation}
Using \eqref{eq:RhoUderivative}, we also rewrite the viscous term as
\begin{equation}\label{eq:momentum_ViscRho}
2\nu\rho\frac{du}{dx'} = 2\rho_{\text{l}}u_{\text{int}}\frac{\nu}{\rho}\frac{d\rho}{dx'}.
\end{equation}

Now, all terms of \eqref{eq:momentum_replacedDerivatives} that were dependent on $u$ are replaced by terms that depend on $\rho$. Hence, we have reached a closed-form equation that depends on one variable only. All terms of the modified momentum conservation equation based on \eqref{eq:momentum_replacedDerivatives}, \eqref{eq:momentum_LeftHandRho}, and \eqref{eq:momentum_ViscRho} can be grouped in the form
\begin{equation} 
\frac{d}{dx'}
\left(
p 
+ \frac{ (\rho_{\text{l}} u_{\text{int}})^2 }{ \rho }
- 2 \rho_{\text{l}} u_{\text{int}} \frac{ \nu }{ \rho }
\frac{ d\rho }{ dx' }
\right) = 0,
\end{equation}
which gives 
\begin{equation} \label{eq:EqPressure}
p 
+ \frac{ (\rho_{\text{l}} u_{\text{int}})^2 }{ \rho }
- 2 \rho_{\text{l}} u_{\text{int}} \frac{ \nu }{ \rho }
\frac{ d\rho }{ dx' } = C_2 = \text{const} .
\end{equation}
Considering that density gradients are equal to zero in the bulk regions, we can evaluate the constant \(C_{2}\) in \eqref{eq:EqPressure} and obtain
\begin{equation} \label{eq:PressureConstant}
C_2=p_{\text{v}}+\frac{\rho_{\text{l}}^2u_{\text{int}}^2}{\rho_{\text{v}}}, \quad \text{and} ~~~ C_2=p_{\text{l}}+\frac{\rho_{\text{l}}^2u_{\text{int}}^2}{\rho_{\text{l}}}.
\end{equation}
Next, we proceed by manipulating the differential equation in a more convenient form to solve.

\subsection{Final ordinary differential equation}

From \eqref{eq:Korteweg}, \eqref{eq:EqPressure} and \eqref{eq:PressureConstant} we have the following ordinary differential equation (ODE) in terms of $x'$:
\begin{subequations}
\begin{equation} \label{eq:PDE}
\kappa \rho \frac{d^2\rho}{dx'^2} 
- \frac{\kappa}{2} \left( \frac{d\rho}{dx'} \right)^2
+ 2\rho_{\text{l}}u_{\text{int}}\frac{\nu}{\rho}\frac{d\rho}{dx'}
= f(\rho), 
\end{equation}
where 
\begin{equation}\label{eq:density_function}
f(\rho) = p_{\text{EOS}}-p_{\text{v}}
-u_{\text{int}}^2\rho_{\text{l}}^2\left(\frac{1}{\rho_{\text{v}}}-\frac{1}{\rho}\right) .
\end{equation}
\end{subequations}
The ODE~\eqref{eq:PDE} can be converted into a more convenient form by applying a simplification based on the assumption that the density function $\rho=\rho(x')$ is monotonically growing, i.e.\ $d\rho/dx'>0$ for all $x'$. Then, the dependency on $x'$ can be transformed into a dependency on $\rho$. 

Considering $d\rho/dx'$ as a function of $\rho$, we have that 
\begin{equation}
\begin{aligned}
\frac{d\rho}{dx'} &= z(\rho), \\
\frac{d^2\rho}{dx'^2} &= \frac{dz}{dx'} \quad\Rightarrow\quad
\frac{d^2\rho}{dx'^2} = \frac{dz}{d\rho}\frac{d\rho}{dx'} \quad\Rightarrow\quad
\frac{d^2\rho}{dx'^2} = \frac{1}{2}\frac{dz^2}{d\rho}.
\end{aligned}
\end{equation}
Also, the first two terms of the left-hand side of \eqref{eq:PDE} can be grouped such that
\begin{equation} \label{eq:transform2}
\begin{aligned} 
& \kappa \rho \frac{d^2\rho}{dx'} 
- \frac{\kappa}{2} \left( \frac{d\rho}{dx'} \right)^2 
= \frac{\kappa\rho}{2}\frac{dz^2}{d\rho}-\frac{\kappa}{2}z^2 \\
&\quad\Rightarrow\quad 
\kappa \rho \frac{d^2\rho}{dx'} 
- \frac{\kappa}{2} \left( \frac{d\rho}{dx'} \right)^2  = 
\frac{\rho^2}{2} \frac{d}{d\rho} \frac{z^2}{\rho}.
\end{aligned}
\end{equation}
By applying the transformations of \eqref{eq:transform2} to \eqref{eq:PDE}, we obtain a non-linear ODE:
\begin{equation} \label{eq:ODE}
\frac{d}{d\rho} \frac{z^2}{\rho} 
+ 4\frac{\rho_{\text{l}}u_{\text{int}}}{\kappa}\frac{\nu}{\rho^3}z
= \frac{2}{\kappa}\frac{f(\rho)}{\rho^2}.
\end{equation}
The ODE~\eqref{eq:ODE} is an Abel's equation of the second kind. 
This equation has the general form\cite{bougoffa2010new}:
\begin{equation} \label{eq:Abel}
[g_0(\rho)+g_1(\rho)z]\frac{dz}{d\rho}=
f_2(\rho)z^2+f_1(\rho)z+f_0(\rho).
\end{equation}
Equation \eqref{eq:ODE} is a particular case of \eqref{eq:Abel} with coefficients:
\begin{equation}
\begin{aligned}
g_0(\rho) = 0, ~&~
g_1(\rho) = \frac{2}{\rho} \\
f_0(\rho) = \frac{2} {\kappa}\frac{f(\rho)}{\rho^2}, ~~
f_1(\rho) =& - 4\frac{\rho_{\text{l}}u_{\text{int}}}{\kappa}\frac{\nu}{\rho^3}, ~~
f_2(\rho) = \frac{1}{\rho^2}.
\end{aligned}
\end{equation}

Looking for a general solution to \eqref{eq:ODE} may not be worth the effort, as the procedure could be too complex for practical applications.
Therefore, we will proceed with an approximate solution. 

\subsection{Exact solution of inviscid case}

The simplest case occurs when there is no viscosity, $\nu=0$. Thus, \eqref{eq:ODE} can be simplified
\begin{equation} \label{eq:ODE_inviscid}
\frac{d}{d\rho}\frac{z^2}{\rho}
= \frac{2}{\kappa}\frac{f(\rho)}{\rho^2}.
\end{equation}
Next, we integrate the whole expression
\begin{equation} \label{eq:Integral_ODE_inviscid}
\frac{z^2(\rho)}{\rho}-\frac{z_{\text{v}}^2}{\rho_{\text{v}}}
= \int_{\rho_{\text{v}}}^{\rho} \frac{2}{\kappa}\frac{f(\rho)}{\rho^2}d\rho,
\end{equation}
where $z_{\text{v}}=z(\rho_{\text{v}})$. The boundary condition $z_{\text{v}}=0$ (no density gradients in the bulk phases) is applied to obtain a solution for $z(\rho)$:
\begin{equation} \label{eq:Solution_inviscid}
z^{(0)}(\rho)
= \sqrt{ \rho\int_{\rho_{\text{v}}}^{\rho} \frac{2}{\kappa}\frac{f(\rho)}{\rho^2}d\rho }.
\end{equation}
The symbol $z^{(0)}$ represents the solution of $z$ for the inviscid case. The selection of the positive sign in the solution \eqref{eq:Solution_inviscid} means that the interface density grows with respect to the $x$ direction. In Figure~\ref{fig:ProblemDescription} this represents the interface in the left-hand side. 

Then, we replace $f(\rho)$ in \eqref{eq:Solution_inviscid} by \eqref{eq:density_function}, and apply the boundary conditions $z_L=z(\rho_L)=0$ (no density gradients in the bulk phases). After this step, we obtain the following expression:
\begin{equation} \label{eq:Integrated_ODE_inviscid}
0 = \int_{\rho_{\text{v}}}^{\rho_{\text{l}}}\frac{p_{\text{EOS}}-p_{\text{v}}}{\rho^2} d\rho
- \int_{\rho_{\text{v}}}^{\rho_{\text{l}}} \frac{u_{\text{int}}^2\rho_{\text{l}}^2}{\rho^2}\left( \frac{1}{\rho_{\text{v}}} - \frac{1}{\rho} \right)d\rho.
\end{equation}
Finally, we perform the integral in the second term of the right-hand side of \eqref{eq:Integrated_ODE_inviscid} and isolate $u_{\text{int}}$:
\begin{equation} \label{eq:Velocity_inviscid}
u_{\text{int}}^{(0)} = \frac{\sqrt{ \int_{\rho_{\text{v}}}^{\rho_{\text{l}}}\frac{p_{\text{EOS}}-p_{\text{v}}}{\rho^2} d\rho }}{ 
\sqrt{ \frac{\rho_{\text{l}}}{\rho_{\text{v}}} \left( \frac{1}{2}\frac{\rho_{\text{l}}}{\rho_{\text{v}}} - 1 \right) + \frac{1}{2} } },
\end{equation}
where $u_{\text{int}}^{(0)}$ represents the interface velocity for the inviscid case. The positive velocity is compatible with the orientation of the interface \eqref{eq:Solution_inviscid} (see Figure~\ref{fig:ProblemDescription}). The inviscid interface velocity is completely independent on $\kappa$. 
An important consideration regarding Eq.\eqref{eq:Velocity_inviscid} is the dependence of the liquid density $\rho_\text{l}$ on the interface velocity. As a result, determining the correct liquid velocity requires solving a coupled system comprising Eq.\eqref{eq:Velocity_inviscid} and the following two additional equations:
\begin{equation} \label{eq:Coupling}
p_\text{l} = p_\text{v} + \rho_\text{l} u_\text{int}^2 \left(
\frac{1}{\rho_\text{v}} 
- \frac{1}{\rho_\text{l}} \right);
\quad p_l = p_{EOS}(\rho_\text{l},T).
\end{equation}
Next, we consider the case with viscous effects.

\subsection{Approximate solution of viscid case}

The equivalent of \eqref{eq:Integral_ODE_inviscid} for the viscid case is:
\begin{equation} \label{eq:Integral_ODE_viscid}
\frac{z^2(\rho)}{\rho} - \frac{z_{\text{v}}^2}{\rho_{\text{v}}}
= \int_{\rho_{\text{v}}}^{\rho} 
\left[
\frac{2}{\kappa}\frac{f(\rho)}{\rho^2}
- 4\frac{\rho_{\text{l}}u_{\text{int}}}{\kappa}\frac{\nu}{\rho^3}
z(\rho)
\right] d\rho.
\end{equation}
The problem with \eqref{eq:Integral_ODE_viscid} is that we cannot use it to explicitly solve for $z$. To overcome this challenge and obtain an approximate solution, we consider that the viscosity $\nu$ has a small influence on the interface profile, which depends on $z$. At least, this approximation will be valid for sufficiently small $\nu$. Thus, we assume that $z$ in the viscous case can be approximated by $z^{(0)}$ from the inviscid case. We use this approximation to compute the integral of the second term on the right-hand side of \eqref{eq:Integral_ODE_viscid}.

Applying the boundary condition $z_{\text{l}}=0$ and the previous assumption to \eqref{eq:Integral_ODE_viscid} yields
\begin{equation} \label{eq:Integrated_viscid}
0
= \int_{\rho_{\text{v}}}^{\rho_{\text{l}}} 
\left[
\frac{2}{\kappa}\frac{f(\rho)}{\rho^2}
- 4\frac{\rho_{\text{l}}u_{\text{int}}}{\kappa}\frac{\nu}{\rho^3}
z^{(0)}(\rho)
\right] d\rho.
\end{equation}
Replacing $f(\rho)$ in \eqref{eq:Solution_inviscid} by \eqref{eq:density_function} leads to a final equation that describes the interface velocity: 
\begin{subequations}
\begin{equation} \label{eq:MainEquation}
A u_{\text{int}}^2 + B u_{\text{int}} + C = 0,
\end{equation}
\begin{equation}
\begin{aligned}
A = \frac{\rho_{\text{l}}}{\rho_{\text{v}}} 
\left( 1 - \frac{1}{2}\frac{\rho_{\text{l}}}{\rho_{\text{v}}} \right) - \frac{1}{2},
\end{aligned}
\end{equation}
\begin{equation} \label{eq:B}
B = - 2 \rho_{\text{l}}
\int_{\rho_{\text{v}}}^{\rho_{\text{l}}}
\frac{\nu}{\rho^3} z^{(0)}
d\rho,
\end{equation}
\begin{equation}
C = \int_{\rho_{\text{v}}}^{\rho_{\text{l}}} \frac{p_{\text{EOS}}-p_{\text{v}}}{\rho^2}d\rho.
\end{equation}
\end{subequations}
Notably, the solution approach above is valid even if $\nu$ depends on $\rho$. 
The interface velocity is obtained by simply solving the quadratic equation, \eqref{eq:MainEquation} toward
\begin{equation} \label{eq:MainSolution}
u_{\text{int}} = \frac{-B-\sqrt{B^2-4AC}}{2A},
\end{equation}
the choice of sign in \eqref{eq:MainSolution} is motivated by the following reasons:
\begin{itemize}
    \item A is always negative ($A<0$) if $\rho_{\text{v}}<\rho_{\text{l}}$, 
    \item B is always negative ($B<0$) if $\rho_{\text{v}}<\rho_{\text{l}}$, 
    \item C is always positive ($C>0$) for $p_{\text{v}}<p_{\text{sat}}(T)$.
\end{itemize}
Consequently, we conclude that a positive solution for $u_i$ is only possible for \eqref{eq:MainSolution}. The positive solution was selected due to the interface orientation choice in \eqref{eq:Solution_inviscid}. The interface velocity must be solved together with the liquid density by coupling \eqref{eq:MainSolution} with \eqref{eq:Coupling}.
The validity of the solution \eqref{eq:MainSolution} is examined in Section~\ref{sec:Results}. 



\section{\label{sec:Numerical}Lattice Boltzmann method and \textit{OpenLB} implementation}

In this work, we compare the analytical results given by \eqref{eq:Velocity_inviscid} and \eqref{eq:MainEquation} with LBM simulations performed using \textit{OpenLB}~\cite{openlb2020}. This is a powerful open-source LBM library applied in various fluid dynamics applications, including sub-grid particulate flows~\cite{bukreev2023simulation}, fully resolved particle flows~\cite{hafen2023simulation,marquardt2024novel}, turbulence simulations~\cite{siodlaczek2021numerical,simonis2022temporal}, optimization~\cite{reinke2022applied,jessberger2022optimization}, sub-grid multiphase flows~\cite{bukreev2023consistent}, and fully resolved multiphase flows~\cite{simonis2024binary}. 
The codes used in this work are implemented in \textit{OpenLB} release 1.8~\cite{olbRelease18} and are available at \url{https://www.openlb.net/download/}.

The basic equation is named the lattice Boltzmann equation~\cite{kruger2017lattice}:
\begin{equation}
f_i(t+\Delta t, \bm{x}+\bm{c}_i\Delta t) 
- f_i(t,\bm{x}) = \frac{\Delta t}{\tau}(f^{\mathrm{eq}}_i-f_i) + \Delta t F_i,
\end{equation}
where $f_i$ and $f_i^{\mathrm{eq}}$ are the distribution function and its equilibrium counterpart. The indexes $i$ indicate the lattice velocity $\bm{c}_i$ in which $f_i$ is evaluated.
The relaxation time $\tau$ is related to the fluid kinematic viscosity $\nu=c_s^2(\tau-0.5\Delta t)$. The parameter $c_s^2=(1/3)(\Delta x/\Delta t)^2$ is called lattice sound speed~\cite{kruger2017lattice}. The equilibrium distribution function is a function of the fluid density $\rho$ and equilibrium velocity $u_{\alpha}^{\mathrm{eq}}$\cite{li2012forcing}:
\begin{equation} \label{eq:equilibrium_function}
f_i^{\mathrm{eq}} = w_i \rho 
\left( 1 + \frac{c_{i\alpha}}{c_s^2}u_{\alpha}^{\mathrm{eq}}
+ \frac{c_{i\alpha}c_{i\beta}-c_s^2\delta_{\alpha\beta}}{2c_s^4}\rho u_{\alpha}^{\mathrm{eq}}u_{\beta}^{\mathrm{eq}} \right),
\end{equation}
where $w_i$ are the lattice weights for each lattice direction $i$. The equilibrium velocity is a quantity used to define $f_i^{\mathrm{eq}}$ and its relation with the real fluid velocity is introduced later.

The term $F_i$ is called forcing scheme and is responsible for the addition of an external force $F_{\alpha}$ to the LBM. Wagner
\cite{wagner2006thermodynamic} proposed the following forcing scheme for the free-energy LBM:
\begin{equation}
F_i = w_i \left( \frac{c_{i\alpha}}{c_s^2}F_{\alpha} 
+ \frac{c_{i\alpha}c_{i\beta}-c_s^2\delta_{\alpha\beta}}{2c_s^4}(F_{\alpha}u_{\beta}^{\mathrm{eq}}+F_{\beta}u_{\alpha}^{\mathrm{eq}} 
+ \psi\delta_{\alpha\beta}) \right),
\end{equation}
with
\begin{equation}
\tau\psi = \left( \tau - \frac{1}{4} \right)
\frac{F_{\alpha}F_{\alpha}}{\rho} + \frac{1}{12}\nabla^2\rho. 
\end{equation}
The Wagner forcing was validated only for 1D cases. Meaning that an extension for 2D cases was not provided yet.
The thermodynamic force is related to the gradient of the chemical potential:
\begin{equation}
F_{\alpha} = - \rho \partial_{\alpha} \mu.
\end{equation}
The spatial derivatives are computed using second order central finite difference stencils. 

The velocity set is the standard two-dimensional nine velocities scheme (D2Q9):
\begin{equation} 
\label{eq:VelocitySet}
\bm{c}_i =
\begin{cases} 
      (0,0), ~~~~~~~~~~~~~~~~~~~~~~~~~~~~~~~~~~~~~~~~~~ i = 0,  \\
      (c,0), (0,c), (-c,0), (0,-c), ~~~~~~ i = 1,...,4, \\
      (c,c), (-c,c), (-c,-c), (c,-c), ~ i = 5,...,8. \\
   \end{cases}
\end{equation}	
The macroscopic variables are computed from the moments of the distribution function:
\begin{subequations}
\begin{equation} \label{eq:Density}
\rho = \sum_i f_i,
\end{equation}	
\begin{equation} \label{eq:Momentum}
\rho u_{\alpha}^{\mathrm{eq}} = \sum_i f_i c_{i\alpha}.
\end{equation}	
\end{subequations}
The real fluid velocity $u_{\alpha}$ depends on $u_{\alpha}^{\mathrm{eq}}$ and the force $F_{\alpha}$:
\begin{equation}
u_{\alpha} = u_{\alpha}^{\mathrm{eq}} + \frac{\Delta t}{2}\frac{F_{\alpha}}{\rho}.
\end{equation}

In this work, we use an equation of state based on the Landau free energy functional \cite{briant2004lattice}:
\begin{subequations}
\begin{equation}
p_{\mathrm{EOS}} = p_c(\nu_{\rho}+1)^2(3\nu_{\rho}^2-2\nu_{\rho}+1-2\tau_w),
\end{equation}
\begin{equation}
\nu_{\rho} = \frac{\rho-\rho_c}{\rho_c}, ~~~
\tau_w = \frac{T_c-T}{T_c},
\end{equation}
\end{subequations}
where $\rho_c$, $p_c$ and $T_c$ are the critical density, pressure, and temperature. The chemical potential is described by:
\begin{equation}
\mu = \frac{4p_c}{\rho_c}\nu_{\rho}
(\nu_{\rho}^2-\tau_w) - \kappa \nabla^2\rho.
\end{equation}

For a planar interface in equilibrium, the bulk densities ($\rho_{\text{v}}^{\text{sat}}$ and $\rho_{\text{l}}^{\text{sat}}$), interface thickness $\xi$, and surface tension $\gamma$ are given by:
\begin{equation}
\begin{aligned}
\rho_{\text{v}}^{\text{sat}} &= \rho_c(1-\tau_w); ~~ 
\rho_{\text{l}}^{\text{sat}} = \rho_c(1+\tau_w); \\ 
\xi &= \sqrt{\frac{\kappa\rho_c^2}{4\tau_wp_c}}; ~~~~
\gamma = \frac{4}{3}\sqrt{2\kappa p_c}(\tau_w)^{3/2}\rho_c.
\end{aligned}
\end{equation}
The superscripts in $\rho_{\text{v}}^{\text{sat}}$ and $\rho_{\text{l}}^{\text{sat}}$ mean that these densities are defined under saturation condition. We define the density ratio under saturation condition:
\begin{equation}
r_{\rho}^{\text{sat}} = 
\frac{\rho_{\text{l}}^{\text{sat}}}{\rho_{\text{v}}^{\text{sat}}}
\quad\Rightarrow\quad
r_{\rho}^{\text{sat}} = \frac{1+\tau_w}{1-\tau_w}.
\end{equation}
The fluid temperature is related to $r_{\rho}^{\text{sat}}$. Thus, throughout this work, we opt to inform the reader $r_{\rho}^{\text{sat}}$ instead of the fluid temperature. 
For instance, our problem has only three independent parameters, which are dimensionless quantities: $p_{\mathrm{v}}/p_{\mathrm{sat}}$, $r_{\rho}^{\text{sat}}$ and $\nu^{\ast}$.

Considering a system of length $L^{\ast}$, we first initialize the density profile using the hyperbolic tangent function:
\begin{equation}
\rho(x^{\ast}) = \rho_{\mathrm{v}}
+ \frac{\rho_{\mathrm{l}}-\rho_{\mathrm{v}}}{2}
\left[ 
\mathrm{tanh} \left( 
\frac{x^{\ast}_1}
{\sqrt{2}\xi^{\ast}} \right)
-
\mathrm{tanh} \left(
\frac{x^{\ast}_2}
{\sqrt{2}\xi^{\ast}} \right)
\right]
\end{equation}
with $x_1^{\ast}=x^{\ast}-0.25L^{\ast}$, $x_2^{\ast}=x^{\ast}-0.75L^{\ast}$ and $\xi^{\ast}=\xi\sqrt{p_{\mathrm{v}}}/(\rho_{\mathrm{v}}\sqrt{\kappa})$.

The initialization of the velocity profile is discussed later. After defining the macroscopic quantities, the distribution function is initialized to be equal to its equilibrium counterpart \eqref{eq:equilibrium_function}. 
Periodic boundary conditions are applied in the bottom and top to reproduce the 1D case. In the side boundaries, the chemical potential and the density are fixed for the vapor value.

There is a key point to discuss regarding viscosity. We have observed that in some simulations, particularly at low viscosity, oscillations in the density and velocity profiles can arise during the initial transient regime. 
To ensure that these oscillations are quickly damped, we assign a relaxation time of 1 in the bulk region while keeping the desired relaxation time at the interface. The higher viscosity in the bulk rapidly suppresses these oscillations, allowing the system to converge more quickly to a steady interface velocity. Importantly, this approach has no impact on the final result. 

\begin{figure}[ht] 
\centering
	\includegraphics[width=\columnwidth]{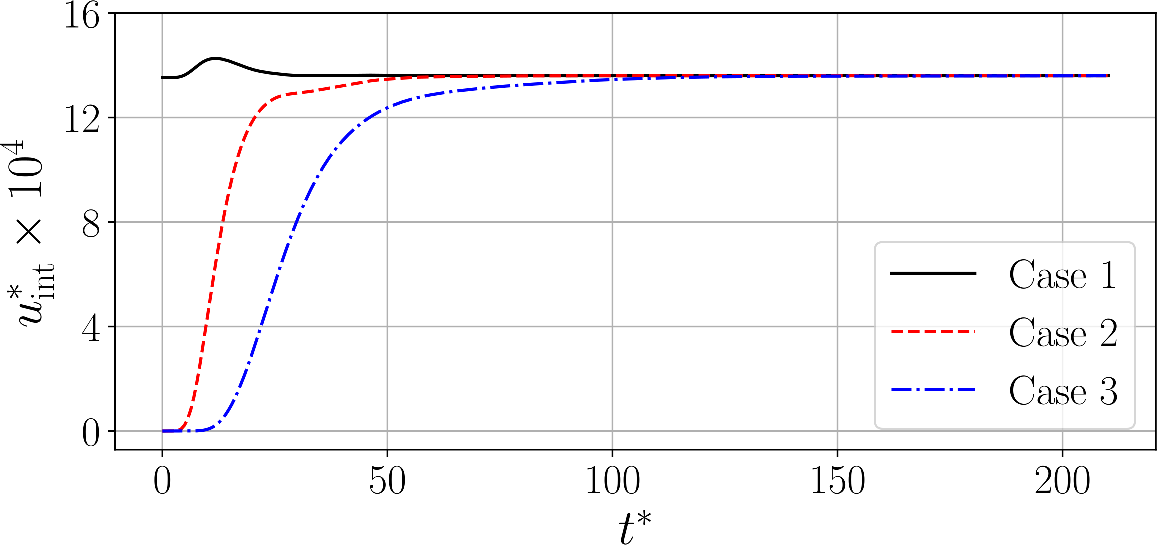}
	\caption{Interface velocity $u_{\mathrm{int}}^{\ast}$ along time $t^{\ast}$ for LBM simulations in \textit{OpenLB}~\cite{openlb2020}. All cases runned with $r_{\rho}^{\text{sat}}=32$, $p_{\mathrm{v}}=0.95p_{\mathrm{sat}}$ and $\nu^{\ast}=1$. Case 1: $L^{\ast}=83$, initial velocity equal to analytical solution \eqref{eq:MainSolution}. Case 2: $L^{\ast}=83$, initial velocity equal to zero. Case 3: $L^{\ast}=166$, initial velocity equal to zero.}
    \label{fig:Configuration_Study}
\end{figure}

The final velocity of the interface should not depend on the initial conditions of the system but rather on the boundary conditions and fluid properties. To verify this, we simulated Cases 1 and 2, as shown in Figure~\ref{fig:Configuration_Study}. In Case 1, the system is initialized with the velocity profile obtained from the analytical solution \eqref{eq:MainSolution} and \eqref{eq:RhoUrelation}. In Case 2, the system is initialized with zero velocity. It is observed that, although the transient period differs between the two cases, the final interface velocity remains entirely independent of the initial condition. Therefore, we chose to initialize the system using the analytical solution to achieve a shorter transient period.

Furthermore, even when initializing with the analytical solution, a slight oscillation in the interface velocity can be observed, as recorded in Figure~\ref{fig:Configuration_Study}. This occurs because $f_i$ is initialized equal to $f_i^{\mathrm{eq}}$, which is only an approximation rather than the exact condition.

As previously mentioned, the analytical solution depends solely on the boundary conditions and fluid properties. Consequently, the final interface velocity should be entirely independent of the domain size $L^{\ast}$. This behavior can be observed in the comparison between Cases 2 and 3 in Figure~\ref{fig:Configuration_Study}. Case 3 shares the same simulation conditions as Case 2 but with a domain size $L^{\ast}$ that is twice as large. Despite differences in the transient regime, the final interface velocity is identical in both cases.

\begin{figure}[ht] 
\centering
	\includegraphics[width=\columnwidth]{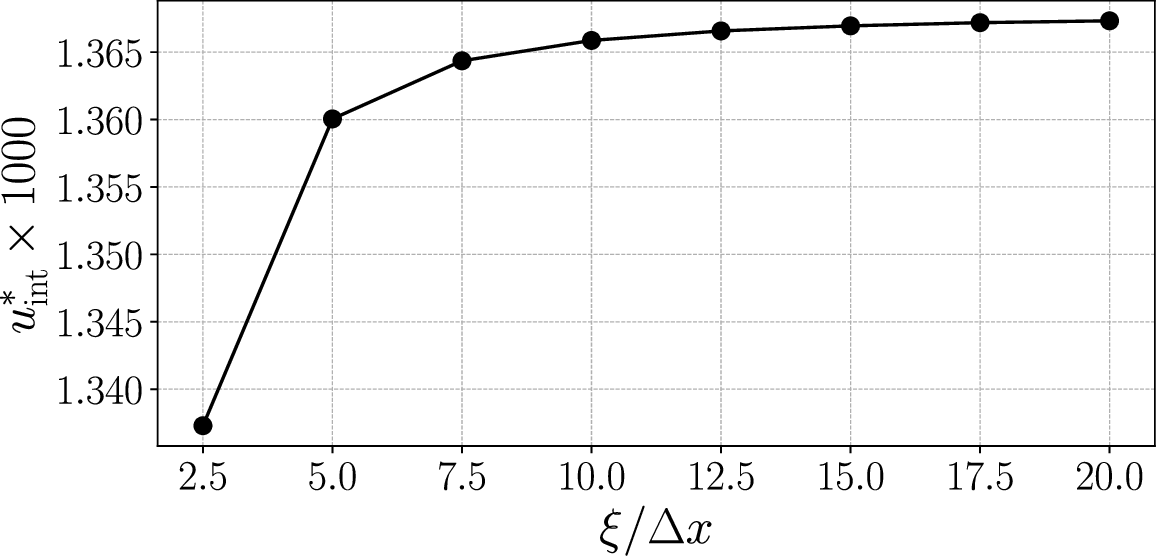}
	\caption{Interface velocity $u_{\mathrm{int}}^{\ast}$ dependency on interface resolution $\xi/\Delta x$ for LBM simulations in \textit{OpenLB}~\cite{openlb2020}. Results at $t^{\ast}=200$ with $L^{\ast}=83$, $r_{\rho}^{\mathrm{sat}}=32$, $p_{\mathrm{v}}=0.95p_{\mathrm{sat}}$ and $\nu^{\ast}=1$.}
    \label{fig:Grid_Study}
\end{figure}

Finally, before presenting the results, we conducted a mesh study to determine the appropriate resolution for the simulations. To this end, we ran a case with $r_{\rho}^{\mathrm{sat}}=32$, $p_{\mathrm{v}}=0.99p_{\mathrm{sat}}$ and $\nu^{\ast}=1$. Since the domain length is not relevant, we used the interface resolution, given by $\xi/\Delta x$, as a measure of the simulation resolution. This value provides an estimate of the number of nodes that occupy the interface.
The results are presented in Figure 4. The difference in $u_{\mathrm{int}}$ between resolutions 10 and 20 was only 0.027\%, indicating that a resolution of 10 is appropriate for simulations.



\section{\label{sec:Results}Results}

\subsection{Verifying modeling validity}

Before testing the inviscid case analytical solution \eqref{eq:Velocity_inviscid} and the viscid case approximate solution \eqref{eq:MainSolution}, we validate the physical equations that lead to our solution. The validations are based on comparisons with numerical simulations using the free-energy LBM proposed by Wagner\cite{wagner2006thermodynamic} and implemented in \textit{OpenLB}\cite{openlb2020}. Based on the results, we show that the effects described here are actually induced by the physical equations. 

Equation~\eqref{eq:EqPressure} is crucial to our work, since it defines the changes of the pressure, given in \eqref{eq:Korteweg}, due to momentum exchange and friction forces acting at the interface. To evaluate the effect of each of those terms, we adopt the following definitions: 

\begin{subequations} \label{eq:pressure_definition}
\begin{equation} \label{eq:non_ideal_pressure}
p_{\text{non-ideal}} \coloneqq p_{\mathrm{EOS}}
+ \frac{\kappa}{2} \left( \frac{ \partial \rho }{ \partial x' } \right)^2
- \kappa \rho \frac{ \partial^2 \rho }{ \partial x'^2 },
\end{equation}
\begin{equation} \label{eq:momentum_pressure}
p_{\text{momentum}} \coloneqq p_{\mathrm{v}} + \rho_{\mathrm{l}}^2u_{\mathrm{int}}^2
\left( \frac{1}{\rho_{\mathrm{v}}} - \frac{1}{\rho} \right) ,
\end{equation}
\begin{equation} \label{eq:friction_pressure}
p_{\text{friction}} \coloneqq p_{\mathrm{v}} + 2\rho_{\mathrm{l}}u_{\mathrm{int}}{\rho}\frac{d\rho}{dx'} ,
\end{equation}
\begin{equation} \label{eq:total_pressure}
p_{\text{total}} \coloneqq p_{\mathrm{v}} + \rho_{\mathrm{l}}^2u_{\mathrm{int}}^2
\left( \frac{1}{\rho_{\mathrm{v}}} - \frac{1}{\rho} \right) 
+ 2\rho_{\mathrm{l}}u_{\mathrm{int}}
\frac{\nu}{\rho}\frac{d\rho}{dx'} .
\end{equation}
\end{subequations}

We perform an LBM simulation using $r_{\rho}^{\mathrm{sat}} = 2$, $p_{\mathrm{v}} = 0.99 p_{\mathrm{sat}}$, and $\nu^{\ast} = 0.5$, running the simulation until a steady-state velocity solution is reached. From the resulting density and velocity fields, we compute \eqref{eq:non_ideal_pressure}, \eqref{eq:momentum_pressure}, \eqref{eq:friction_pressure}, and \eqref{eq:total_pressure}. The first- and second-order derivatives in these equations are approximated using central finite differences with second-order accuracy.

\begin{figure}[ht] 
\centering
	\includegraphics[width=\columnwidth]{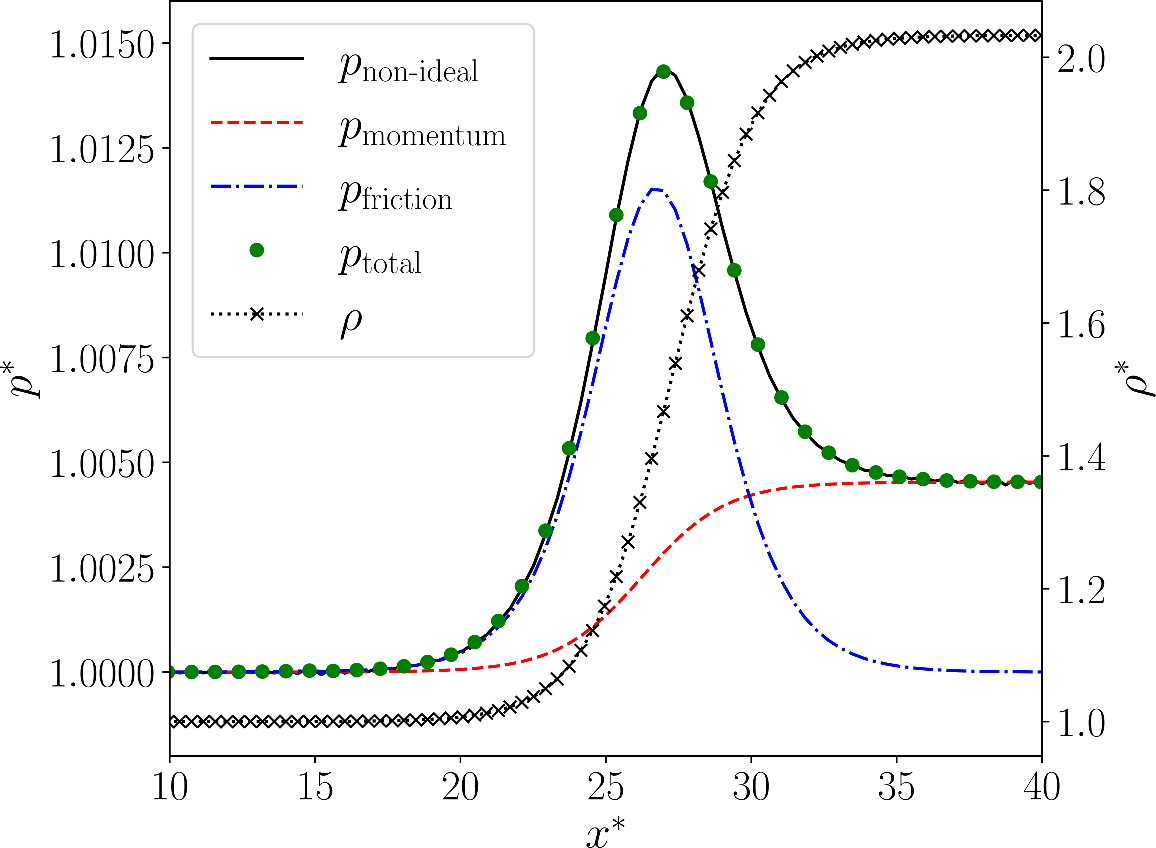}
	\caption{Pressure $p^{\ast}$ (left vertical axis) and density $\rho^{\ast}$ (right vertical axis) along domain length $x^{\ast}$. Dimensionless quantities are defined in \eqref{eq:dimensionless}. Pressure definitions are given in \eqref{eq:pressure_definition}. Simulation results with $r_{\rho}^{\mathrm{sat}}=2$, $p_{\mathrm{v}}=0.99p_{\mathrm{sat}}$ and $\nu^{\ast}=1$ obtained from LBM implemented in \textit{OpenLB}\cite{openlb2020}.}
    \label{fig:ModelValidity}
\end{figure}

In Figure~\ref{fig:ModelValidity} we show the density profile for this simulation and how $p_{\text{non-ideal}}$ and $p_{\text{total}}$ change along the interface region. Both pressures are equivalent which is consistent with \eqref{eq:EqPressure}. We observe that going from the vapor region and crossing the interface, $p_{\text{non-ideal}}$ grows, reaching a peak inside the interface region and then decreasing towards the liquid region. The value of $p_{\text{non-ideal}}$ is larger inside the liquid region in comparison with the vapor region.

In Figure~\ref{fig:ModelValidity} we observe that $p_{\text{momentum}}$ grows continuously from the vapor region to the liquid region due to momentum exchange across the interface. In the bulk regions $p_{\text{non-ideal}}$ is equal to $p_{\text{momentum}}$ indicating that the momentum exchange is the only responsible for the pressure difference between bulk phases.

Finally, we also plot $p_{\text{friction}}$ in Figure~\ref{fig:ModelValidity}. The value of this quantity grows inside the interface due to the velocity gradient across the interface. However, the friction does not contribute to change the pressure between the two bulk phases. The numerical results support our physical model. Any possible physical inconsistency of this solution will be related to the diffuse interface model and not with any further consideration.

\subsection{Comparison with lattice Boltzmann method}

We compare the analytical approximation given by \eqref{eq:MainSolution} against results obtained from LBM simulations. To assess the model’s robustness, we consider three representative density ratios, $r_{\rho}^\text{sat} = 2$, $8$, and $32$. We then analyze how the interface velocity $u_{\mathrm{int}}^{\ast}$ varies as a function of the kinematic viscosity $\nu^{\ast}$. Simulations are performed across a range of pressure ratios defined as $f_\text{p} = p_\text{v}/p_\text{sat}$.

It is important to note that the viscid solution given by \eqref{eq:MainSolution} is an approximation and may deviate from the exact solution of the underlying differential equation. However, the free-energy LBM is a consistent numerical method for solving the mass and momentum conservation equations, and its results converge to the exact solution as the spatial and temporal resolution increases. Therefore, we use high-resolution LBM simulations as a reference to assess the accuracy of our analytical approximation.

\begin{figure}[ht] 
\centering
	\includegraphics[width=\columnwidth]{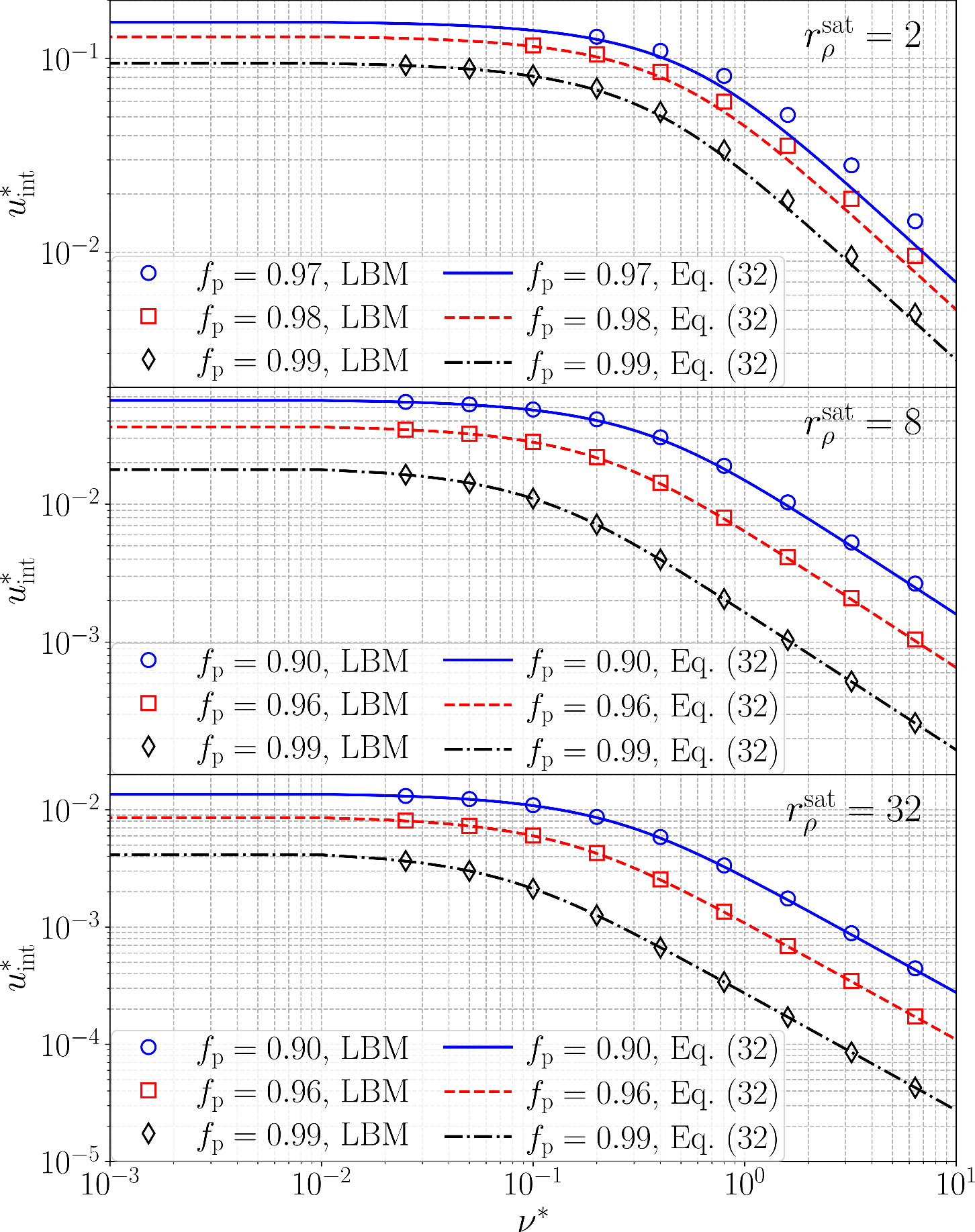}
	\caption{Dependency of interface velocity $u^{\ast}_{\mathrm{int}}$ on fluid kinematic viscosity $\nu^{\ast}$ for fixed $r_\rho^{\text{sat}}$ and different values of $f_\text{p}=p_\mathrm{v}/p_\text{sat}$. Dimensionless quantities are defined in \eqref{eq:dimensionless}. 
    Points represent simulation results using LBM implemented in \textit{OpenLB}\cite{openlb2020}. Lines represent approximate solution \eqref{eq:MainSolution}.}
    \label{fig:rTotal}
\end{figure}

The results of this comparison are shown in Figure~\ref{fig:rTotal}.
The overall behavior of the solution is qualitatively similar across all cases. For very low viscosities, the solution converges to the inviscid limit given by \eqref{eq:Velocity_inviscid}. Conversely, when $\nu^\ast$ is sufficiently large, the interface velocity becomes small and the quadratic term $u_\text{int}^2$ in \eqref{eq:MainEquation} can be neglected. In this high-viscosity regime, the interface velocity exhibits an inverse dependence on the viscosity.

A selection of relative errors between the analytical approximation and the LBM results is compiled in Table~\ref{tab:errors}. Three main trends can be identified from the data:
First, the approximation becomes less accurate as the viscosity $\nu^\ast$ increases. This is consistent with the construction of our solution, which is derived from the inviscid limit and, therefore, performs best in low-viscosity regimes.
Second, the error also increases as the pressure fraction $f_\text{p} = p_\text{v} / p_\text{sat}$ decreases. 
A third trend is observed with respect to the density ratio. For lower values such as $r_\rho^\text{sat} = 2$, the approximation becomes significantly less accurate. This suggests that, near the critical point—where the density ratio approaches unity—the interface becomes more sensitive to viscous effects. Since our approximation relies on the assumption of proximity to the inviscid case, its accuracy degrades in this regime, where the influence of viscosity becomes more pronounced.

\begin{table}[htbp]
\centering
\caption{Relative error between the analytical approximation and LBM simulations for different density ratios $r_\rho^\text{sat}$, pressure fractions $f_\text{p}$, and viscosities $\nu^\ast$.}
\label{tab:errors}
\begin{tabular}{c@{\hspace{10pt}}c@{\hspace{10pt}}c@{\hspace{10pt}}c}
\hline
$r_\rho^\text{sat}$ & $f_\text{p}$ & $\nu^\ast$ & Relative error (\%) \\
\hline
32 & 0.99 & 0.025 & 0.02 \\
32 & 0.99 & 6.4   & 0.28 \\
32 & 0.90 & 0.025   & 0.35 \\
32 & 0.90 & 6.4   & 2.91 \\
8  & 0.99 & 0.025   & 0.024 \\
8  & 0.99 & 6.4   & 0.26 \\
8  & 0.90 & 0.025   & 0.10 \\
8  & 0.90 & 6.4   & 3.10 \\
2  & 0.99 & 0.025    & 0.20 \\
2  & 0.99 & 6.4    & 10.02 \\
2  & 0.97 & 0.2    & 3.10 \\
2  & 0.97 & 6.4    & 24.81 \\
\hline
\end{tabular}
\end{table}

These observations help delineate the validity range of the proposed approximation. The solution consistently yields accurate results in the low-viscosity regime. Moreover, we observe that if the density ratio remains above 8 and the pressure fraction exceeds 0.90, the approximation remains reliable across a broad range of viscosities. For $r_\rho^\text{sat} = 2$, however, some low-viscosity cases could not be simulated due to stability limitations inherent to the BGK collision operator at small relaxation times.

\subsection{Comparison with sharp interface solution}

Jamet\cite{jamet2004test} modeled a similar problem using a sharp-interface approach. His formulation is based on the framework originally proposed by Ishii\cite{ishii1975thermo}, which derives from applying jump conditions across the interface under the assumption of thermodynamic equilibrium.
Under this assumption, the following equation is obtained (see Ishii\cite[p.41, Eq.(2-106)]{ishii1975thermo}; Jamet\cite[Eq.~(16)]{jamet2004test}):
\begin{equation}
p_\text{v} = p_\text{sat} - \frac{1}{2}\frac{\rho_v\rho_l}{\rho_l - \rho_v} (u_v - u_l)^2.
\end{equation}
Assuming symmetry in our test case, such that $u_\text{l} = 0$, and using \eqref{eq:RhoUrelation}, we obtain:
\begin{equation} \label{eq:Jamet}
u_\text{int} = \sqrt{ \frac{p_\text{sat} - p_\text{v}}
{\frac{\rho_\text{l}^2}{2} \left( \frac{1}{\rho_\text{v}} - \frac{1}{\rho_\text{l}} \right)} }.
\end{equation}

This relation shares similarities with \eqref{eq:Velocity_inviscid}. In fact, \eqref{eq:Jamet} can be derived from \eqref{eq:Velocity_inviscid} if we assume:
\begin{equation}
\int_{\rho_v}^{\rho_l} \frac{p_\text{EOS}}{\rho^2} \, d\rho
= \int_{\rho_v}^{\rho_l} \frac{p_\text{sat}}{\rho^2} \, d\rho.
\end{equation}
This is essentially the Maxwell construction, which holds under thermodynamic equilibrium. However, under non-equilibrium conditions, this approximation may no longer be valid. Moreover, this macroscopic model does not capture the dependence of the evaporation flux on viscosity.

\begin{figure}[ht] 
\centering
	\includegraphics[width=\columnwidth]{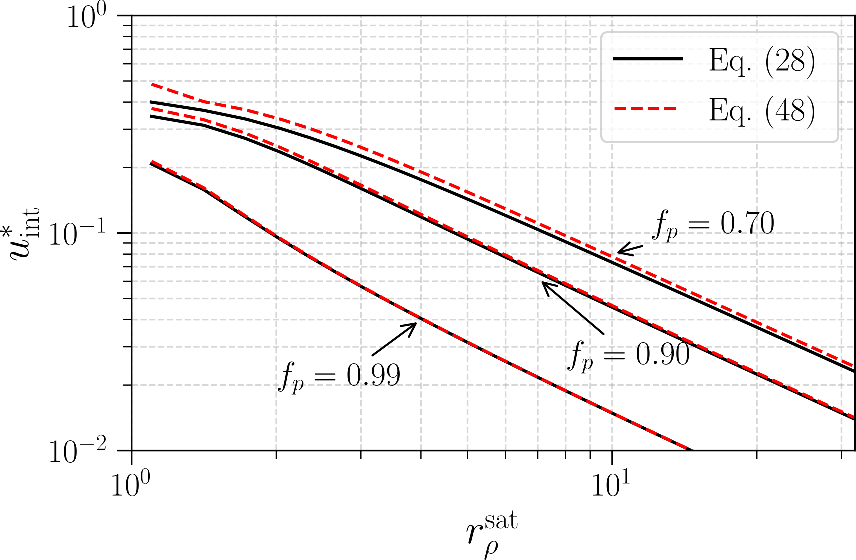}
	\caption{Dependency of interface velocity $u^{\ast}_{\mathrm{int}}$ on the density ratio $r_\rho^\text{sat}$ for different values of $f_\text{p}=p_\mathrm{v}/p_\text{sat}$. Plot shows a comparison between our inviscid solution \eqref{eq:Velocity_inviscid} and Jamet solution \eqref{eq:Jamet}.}
    \label{fig:Sharp}
\end{figure}

Next, we compare our inviscid solution \eqref{eq:Velocity_inviscid} with the sharp-interface model given by \eqref{eq:Jamet}.
The results for different values of $r_\rho^\text{sat}$ and $f_\text{p}$ are shown in Figure~\ref{fig:Sharp}.
Overall, the two solutions exhibit excellent agreement. Noticeable deviations are observed only when $f_\text{p}$ is significantly reduced (e.g., to 0.7), or near the critical point, where $r_\rho^\text{sat}$ approaches unity.

For general purposes, we conclude that the two models yield comparable predictions across a wide range of conditions.
In future work, we intend to extend this comparison by incorporating LBM simulations using more stable collision operators.
At present, the use of the BGK operator restricts us from accessing sufficiently low viscosity to fully explore the inviscid regime.



\section{\label{sec:Conclusion}Conclusion}

This work presents an analytical approximation to the problem of isothermal evaporation under sub-saturation pressure conditions using a diffuse interface model. For the inviscid case, our solution is exact. By avoiding traditional assumptions such as local thermodynamic equilibrium at the interface, the proposed approach offers a new perspective on the relationship between evaporation rates and viscosity. The derived solution is evidenced against numerical simulation results produced with an LBM implemented in \textit{OpenLB}, demonstrating excellent agreement and reinforcing its physical consistency.

The findings highlight the potential of diffuse interface models in studying phase change phenomena, not only as a predictive tool but also as a benchmark for numerical methods. The results provide clear evidence that the evaporation dynamics, including interface velocity, are strongly influenced by fluid viscosity, which is often overlooked in existing literature. Furthermore, this found analytical solution can serve as a reference for experimental validation of diffuse interface methodologies or as a basis for future refinements of these models.

Future work could extend this approach to non-isothermal conditions and multi-component systems, enabling broader applications in practical scenarios such as fuel atomization and heat transfer processes. By bridging the gap between analytical models and numerical simulations, this study contributes to a more comprehensive and accurate modeling of phase-change phenomena in complex systems.





\section*{Acknowledgments}

We acknowledge the support of the Alexander von Humboldt Foundation in sponsoring the postdoctoral fellowship of researcher Dr.\ Luiz Eduardo Czelusniak at LBRG, KIT.








\providecommand{\noopsort}[1]{}\providecommand{\singleletter}[1]{#1}%

\end{document}